\documentclass[conference,a4paper]{IEEEtran}

\usepackage{xcolor}
\usepackage{balance}
\usepackage{cite}
\usepackage{multirow}
\usepackage[pdftex]{graphicx}
\usepackage{amsmath,amssymb,amsfonts}
\usepackage{textcomp}
\usepackage[caption=false,font=footnotesize]{subfig}
\usepackage[nolist]{acronym}
\usepackage[capitalize]{cleveref}
\usepackage{nicefrac}
\usepackage{graphicx}  % remove 'demo' option for your real document
\usepackage{amsmath}
\newcommand\norm[1]{\left\lVert#1\right\rVert}

\crefname{table}{Tab.}{Tabs.}
\Crefname{table}{Tab.}{Tabs.}
\crefformat{equation}{(#2#1#3)}

\begin{document}
\begin{acronym}
\acro{AWG}[AWG]{Arbitrary Waveform Generator}
\acro{CW}[CW]{Continuous Wave}
\acro{HST}[HST]{High-Speed Train}
\acro{IF}[IF]{Intermediate Frequency}
\acro{LO}[LO]{Local Oscillator}
\acro{LSF}[LSF]{local scattering function}
\acro{MMW}[mmWave]{millimeter wave}
\acro{OFDM}[OFDM]{orthogonal frequency-division multiplexing}
\acro{PCB}[PCB]{Printed Circuit Board}
\acro{SMD}[SMD]{Surface Mount Device}
\acro{SNR}[SNR]{signal-to-noise ratio}
\acro{RF}[RF]{radio frequency}
\acro{V2X}[V2X]{vehicle-to-everything}
\acro{IC}[IC]{Integrated Circuit}
\acro{FPGA}[FPGA]{Field Programmable Gate Array}
\acro{ISI}[ISI]{Inter-Symbol Interference}
\acro{CSIT}{channel state information at the transmitter}
\acro{CSI}{channel state information}
\acro{ML}{Machine Learning}
\acro{RNN}{Recurrent Neural Networks}
\acro{LSTM}{Long Short-Term Memory}
\acro{GRU}{Gated Recurrent Unit}
\acro{FDD}{frequency-division duplex}
\acro{TDD}{time-division duplex}
\acro{CIR}{channel impulse response}
\acro{CTF}{channel transfer function}
\acro{PDP}{power delay profile}
\acro{DSD}{Doppler power spectral density}
\acro{IFFT}{Inverse Fast Fourier Transform}
\acro{ITS}[ITS]{intelligent transportation systems}
\acro{5G}[5G]{fifth generation}
\acro{NR}[NR]{new radio}
\acro{QAM}[QAM]{quadrature amplitude modulation}
\acro{ICI}[ICI]{Inter-Carrier Interference}
\acro{MSE}[MSE]{mean square error}
\acro{BER}[BER]{bit error ratio}
\acro{RMS}[RMS]{root-mean-square}
\acro{TDL-A}[TDL-A]{tapped delay line A}
\acro{ISI}[ISI]{Inter-Symbol Interference}
\acro{DFT}[DFT]{discrete Fourier transform}
\acro{IDFT}[IDFT]{inverse discrete Fourier transform}
\acro{CTF}[CTF]{channel transfer function}
\acro{DR}[DR]{dynamic range}
\acro{HPBW}[HPBW]{half-power beamwidth}
\acro{DPSS}[DPSS]{discrete prolate spheroidal sequences}
\acro{CDF}[CDF]{cumulative distribution function}
\acro{URLLC}[URLLC]{Ultra-Reliable Low-Latency Communication}
\acro{3GPP}[3GPP]{3rd Generation Partnership Project}
\acro{MIMO}[MIMO]{multiple-input multiple-output}
\acro{SISO}[SISO]{single-input single-output}
\acro{MRC}[MRC]{maximal ratio combining}
\acro{AWGN}[AWGN]{additive white Gaussian noise}
\acro{LS}[LS]{least-square}
\acro{SVD}[SVD]{singular value decomposition}
\acro{SE}[SE]{spectral efficiency}
\acro{ULA}[ULA]{uniform linear array}
\acro{FSPL}[FSPL]{free space path loss}
\acro{LOS}[LOS]{line-of-sight}
\end{acronym}
%\title{Methods to Estimate Millimeter Wave Channels using Sub-6\,GHz Out-of-Band Information}
%\title{Estimating mmWave MIMO Channels using sub-6\,GHz Out-of-Band Information}
\title{Channel Estimation for mmWave MIMO \\ using sub-6\,GHz Out-of-Band Information}

%\title{Sub-6\,GHz Out-of-Band Information Assisted Channel Estimation for Millimeter Wave Systems}
\author{\IEEEauthorblockN{
Faruk Pasic\IEEEauthorrefmark{1},
Markus Hofer\IEEEauthorrefmark{3},
Mariam Mussbah\IEEEauthorrefmark{1}\IEEEauthorrefmark{2},
Sebastian Caban\IEEEauthorrefmark{1},
Stefan Schwarz\IEEEauthorrefmark{1}, \\
Thomas Zemen\IEEEauthorrefmark{3} and
Christoph F. Mecklenbr{\"a}uker\IEEEauthorrefmark{1}
}%

\IEEEauthorblockA{\IEEEauthorrefmark{1}% 2nd affiliations
Institute of Telecommunications, TU Wien, Vienna, Austria}
\IEEEauthorblockA{\IEEEauthorrefmark{2}% 2nd affiliations
Christian Doppler Laboratory for Digital Twin assisted AI for sustainable Radio Access Networks}
\IEEEauthorblockA{\IEEEauthorrefmark{3}% 3rd affiliations
AIT Austrian Institute of Technology, Vienna, Austria} 
\IEEEauthorblockA{faruk.pasic@tuwien.ac.at}
}

\IEEEoverridecommandlockouts
\IEEEpubid{\makebox[\columnwidth]{979-8-3503-8532-8/24/\$31.00~\copyright2024 IEEE \hfill} \hspace{\columnsep}\makebox[\columnwidth]{ }}

%979-8-3503-8532-8/24/$31.00 ©2024 IEEE

\maketitle
\IEEEpubidadjcol

\begin{abstract}
Future wireless \ac{MIMO} communication systems will employ sub-6\,GHz and \ac{MMW} frequency bands working cooperatively.
%Configuring these \ac{MMW} antenna arrays can be very challenging.
Establishing a \ac{MIMO} communication link usually relies on estimating \ac{CSI} which is difficult to acquire at \ac{MMW} frequencies due to a low \ac{SNR}.
In this paper, we propose three novel methods to estimate \ac{MMW} \ac{MIMO} channels using out-of-band information obtained from the sub-6\,GHz band.
We compare the proposed channel estimation methods with a conventional one utilizing only in-band information.
Simulation results show that the proposed methods outperform the conventional \ac{MMW} channel estimation method in terms of achievable spectral efficiency, especially at low \ac{SNR} and high K-factor. 
% The estimation accuracy and spectral efficiency can be further improved by utilizing out-of-band infromation.
% It is verified by simulation results that the proposed methods significantly outperform .... in spectral efficiency. 

\end{abstract}
\vskip0.5\baselineskip

\acresetall

% Plan
% 1. Motivate the use of the higher frequency bands (sub 6 GHz -> mmWave)
% 2. Distadvantages of mmWave: large PL, low SNR -> inacurrate CSI
% 3. Through MB measurements shown that channels have similar characteristics
% 4. So far, methods to use OOB info are... but channel estimation missing
% 5. Contribution
% 6. Organization
% 7. Notation

\section{Introduction}
Today’s wireless communication systems operate mainly in the sub-6\,GHz frequency bands. 
Due to the spectrum shortage, sub-6\,GHz bands can not keep up with the growing demand for high data rates.
Fortunately, significantly more bandwidth is available in \ac{MMW} bands (10\,GHz -- 300\,GHz), which enables high data rate transmissions~\cite{Ai2020}.
%When combined with \ac{MIMO} systems, \ac{MMW} communication can achieve high data rates and therefore is considered as a promising technology for next-generation wireless systems~\cite{Rappaport2013}
%To provide sufficient link margin, \ac{MMW} systems employ \ac{MIMO} technology along with large antenna arrays and directional beamforming~\cite{Pi2011}.
To provide a sufficient link margin, in most \ac{MMW} systems, antenna arrays will be used at the transmitter and receiver side.
This creates many opportunities to apply \ac{MIMO} communication techniques such as directional beamforming~\cite{Heath2016}.
Therefore, \ac{MMW} \ac{MIMO} communication is a promising technology for next-generation wireless systems~\cite{Rappaport2013}.

Configuring \ac{MMW} antenna arrays is challenging.
The main challenge in using \ac{MMW} frequency bands is link establishment, which is done by designing the precoder and combiner~\cite{Nuria2017}.
The precoder and combiner design usually relies on \ac{CSI}, which is difficult to acquire at \ac{MMW} frequencies due to low pre-beamforming \ac{SNR}.
Hence, the performance of \ac{CSI} estimation is directly related to the coverage capability of wireless systems, which is the main obstacle of \ac{5G} and beyond systems employing \ac{MMW} bands.

Many multi-band measurement campaigns in different environments have shown that \ac{MMW} frequency bands have similar propagation characteristics as sub-6\,GHz bands~\cite{Hofer2021, Hofer2022, Dupleich2019, Pasic2022, Pasic2023, Pasic2023_mag}.
Specifically, the multipath components were found to be similarly distributed in these two frequency bands.
Since \ac{MMW} systems will likely be deployed in conjunction with sub-6\,GHz systems~\cite{Kishiyama2013}, the opportunity arises to use these two bands cooperatively to achieve a better system throughput.

So far, several beam-selection strategies to leverage sub-6\,GHz out-of-band information as side information on \ac{MMW} band have been proposed.
In~\cite{Ali2018}, the authors propose using spatial information extracted at sub-6\,GHz to improve \ac{MMW} compressed beam selection.
Furthermore, two approaches (non-parametric and parametric) to translate the lower frequency spatial correlation to the higher frequency have been proposed in~\cite{Ali2016}.
%In~\cite{Ali2019}, authors use the sub-6\,GHz channel covariance as an out-of-band side information for \ac{MMW} compressed covariance estimation.
In~\cite{Kyosti2023}, the authors analyze the feasibility of using low-band channel information for coarse estimation of high-band beam directions.
The authors in~\cite{Sim2020, Ma2023} propose a deep learning-based beam selection algorithm for \ac{MMW} bands, exploiting sub-6\,GHz channel information.
However, the idea of exploiting sub-6\,GHz out-of-band information for \ac{MMW} channel estimation has not been investigated yet.

\textbf{Contribution:}
In this paper, we propose three novel pilot-aided channel estimation methods for \ac{MMW} \ac{MIMO} systems based on out-of-band information.
We exploit the relationship between \ac{LOS} channel components across different frequency bands and estimate the \ac{MMW} channel with the aid of adapted channel coefficients obtained in the sub-6\,GHz band.
We evaluate the proposed methods through simulations in terms of \ac{SE}.

%In this paper, we propose three pilot-aided channel estimation methods for \ac{MMW} \ac{MIMO} systems based on side information obtained in the sub-6\,GHz band.
%We exploit similar propagation characteristics between signals in different frequency bands and estimate \ac{MMW} channel with the aid of side information obtained in sub-6\,GHz band.
%We exploit similar propagation characteristics between signals in different frequency bands and estimate \ac{MMW} channel with the aid of channel coefficients obtained in sub-6\,GHz band.
%Thereby, we reduce the overhead of configuring \ac{MMW} link and increase system throughput.
%We analyze the impact of inter-band channel correlation on the proposed methods.
% We investigate the channel correlation over different frequency bands for different scenarios. 
%Further, we evaluate the proposed channel estimation methods by comparing them with ones that do not use out-of-band information through link-level simulations in terms of channel \ac{MSE} and achievable throughput.

%The rest of the paper is organized as follows. 
\textbf{Organization:}
In~\cref{sec:system_model}, we present the system model considered in this work.
\cref{sec:methods} presents the proposed channel estimation methods.
The performance comparison is given in~\cref{sec:comparison}.
Finally,~\cref{sec:conclusion} concludes the paper.

\textbf{Notation:} Bold uppercase letters ${\mathbf X}$ denote matrices  and bold lowercase letters ${\mathbf x}$ denote vectors.
We use the superscript $\left( \cdot \right) ^{\rm H}$ for Hermitian transposition and the superscript $\left( \cdot \right) ^{\left( \rm b \right)}$ for frequency-band dependent values, where ${\rm b} \in \{ {\rm s}, {\rm m} \}$.
Here, ${\rm s}$ denotes the sub-6\,GHz frequency band and ${\rm m}$ denotes the \ac{MMW} frequency band.
Operation $\odot$ represents element-wise multiplication and $\lVert \cdot \rVert_F$ denotes the Frobenius norm.

\section{System Model} \label{sec:system_model}
\begin{figure}[t]
    \centering
    {\includegraphics[width=\columnwidth]{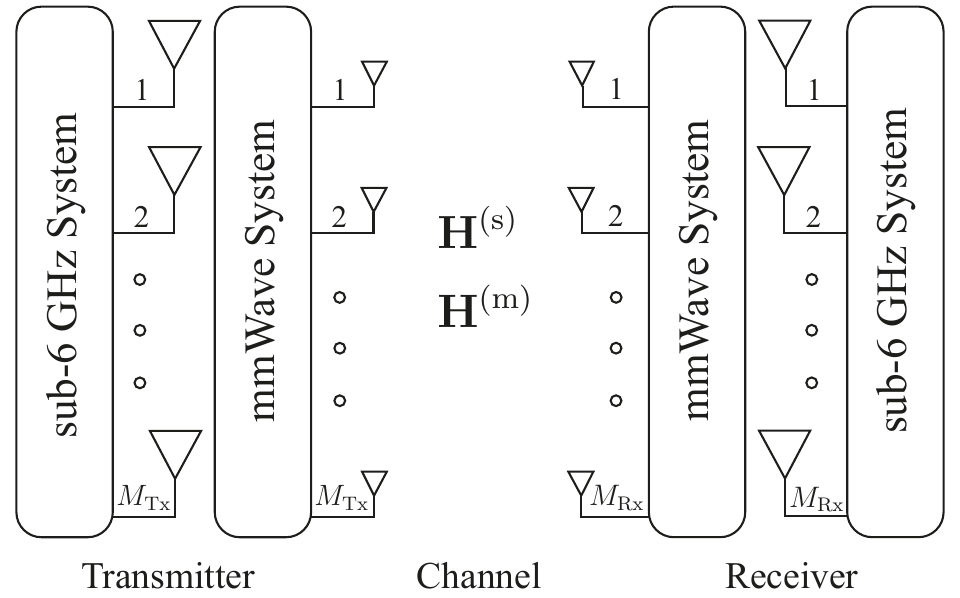}}
    \caption{A multi-band \ac{MIMO} transmission system with co-located sub-6 \,GHz and \ac{MMW} antenna arrays. $\mathbf{H}^{\left( \rm s \right)}$ denotes the sub-6\,GHz \ac{MIMO} channel. $\mathbf{H}^{\left( \rm m \right)}$ denotes the \ac{MMW} \ac{MIMO} channel.}
    \label{fig:system_model}
\end{figure}

We consider a point-to-point multi-band \ac{MIMO} system, where sub-6\,GHz and \ac{MMW} systems operate simultaneously.
The transmitter consists of $M_{\rm Tx}$ sub-6\,GHz and $M_{\rm Tx}$ \ac{MMW} antenna elements, while the receiver is equipped with an antenna array comprising $M_{\rm Rx}$ sub-6\,GHz and $M_{\rm Rx}$ \ac{MMW} antenna elements.
%Both transmitters at sub-6\,GHz and \ac{MMW} consist of $M_{\rm Tx}^{\left( \rm s \right)} = M_{\rm Tx}^{\left( \rm m \right)}$ antenna elements, while the receivers are equipped with $M_{\rm Rx}^{\left( \rm s \right)} = M_{\rm Rx}^{\left( \rm m \right)}$ antenna elements.
The sub-6\,GHz and \ac{MMW} parts of the system are arranged as a \ac{ULA} of dipole antennas, modeled as isotropic point sources (see~\cref{fig:system_model}).
We assume that the sub-6\,GHz and \ac{MMW} antenna arrays are co-located and aligned. 
Both arrays have the same number of antenna elements mutually separated by 0.5$\,\lambda^{\left( \rm m \right)}$, where $\lambda^{\left( \rm m \right)}$ represents the wavelength of the \ac{MMW} system.
The small antenna spacing for the sub-6\,GHz array can be achieved by a compact design, as proposed in~\cite{Getu2005}.
Furthermore, we assume perfect time and frequency synchronization at the receiver.
The transmitter and receiver are equipped with one \ac{RF} chain per antenna, thereby allowing for fully digital beamforming for both sub-6\,GHz and \ac{MMW} systems~\cite{Ghods2019}. 

%We assume a narrowband signal model for both the sub-6 GHz and the mm Wave systems.
%Further, the coherence time of the channel is long enough to permit (i) retrieving angular information at sub-6 GHz, and (ii) using it for mmWave CBS. 
%This assumption is reasonable with directional beamforming at mmWave [12]. 

We consider an \ac{OFDM} system with $N^{\left( \rm b \right)}$ subcarriers.
Blocks of $N^{\left( \rm b \right)}$ symbols modulated by \ac{QAM} are mapped onto $N^{\left( \rm b \right)}$ different subcarriers to construct \ac{OFDM} symbols.
%The signal to be transmitted from each antenna is subject to the \ac{OFDM} procedures of \ac{IDFT} and cyclic prefix insertion and then, at the receiver, removal of cyclic prefix followed by \ac{DFT}.
The \ac{OFDM} system converts a broadband frequency-selective channel into narrowband frequency flat channels with the help of a \ac{DFT} and application of a cyclic prefix~\cite{Cimini1985,Schwarz2010}. 
The channel at the \ac{OFDM} subcarrier $n$ is described by an $M_{\rm Rx} \times M_{\rm Tx}$ dimensional complex-valued channel matrix $\mathbf{H}^{\left( \rm b \right)} [n]$, employing the equivalent complex baseband representation of the \ac{OFDM} system.
%The channel at \ac{OFDM} subcarrier $k$ and symbol-time $n$ is described by $M_{\rm Rx} \times M_{\rm Tx}$ dimensional complex-valued channel matrix $\mathbf{H}^{\rm b} [k,n]$, employing the equivalent complex baseband representation of the \ac{OFDM} system.

\subsection{Channel Model} \label{sec:channel_model}
% Foundations of MIMO - Book
%The simplest possible analytical model, sometimes referred to as the canonical channel, is the one where the entries of H are i.i.d.
%The i.i.d. Rayleigh-faded model is typically justified as representative of environments with rich scattering. 
%This is a simplification of reality, since simultaneous and complete decorrelation of more than two antennas may be geometrically impossible, but the model has enormous value nevertheless; it enables exploring and demarcating what may be feasible, and in that respect it played a very important role in the early days of \ac{MIMO}.

We consider a frequency-selective Rician fading channel model defined as
\begin{equation}
    \mathbf{H}^{\left( \rm b \right)} [n] = 
     \sqrt{\eta^{\left( \rm b \right)}} \left( \mathrm{A_{\rm fs}^{\left( \rm b \right)}} \mathbf{H}_{\rm fs}^{\left( \rm b \right)} [n] +
    \mathrm{A_{\rm rp}^{\left( \rm b \right)}} \mathbf{H}_{\rm rp}^{\left( \rm b \right)} [n] \right)
    \label{eq:channel_model}
\end{equation}
where $\eta^{\left( \rm b \right)}$ is the path loss coefficient,  $\mathrm{A_{\rm fs}^{\left( \rm b \right)}} = \sqrt{K^{\left( \rm b \right)} / \left( 1+K^{\left( \rm b \right)} \right)}$ denotes the free-space scaling factor, $\mathrm{A_{\rm rp}^{\left( \rm b \right)}} = \sqrt{1 / \left( 1+K^{\left( \rm b \right)} \right) }$ represents the Rayleigh-part scaling factor and $K^{\left( \rm b \right)}$ denotes the Rician $K$-factor.
Based on the measurements presented in~\cite{Miao2023}, due to the less pronounced multipath components at \ac{MMW} bands compared to sub-6\,GHz bands, we assume that the \ac{MMW} $K$-factor $K^{\left( \rm m \right)}= 10 K^{\left( \rm s \right)}$. % is 10 times larger that the sub-6\,GHz $K$-factor $K^{\left( \rm s \right)}$.
%The $K$-factor is assumed to be identical for sub-6\,GHz and \ac{MMW} frequency bands.
The deterministic free-space channel $\mathbf{H}_{\rm fs}^{\left( \rm b \right)} [n]$  is defined by
\begin{equation}
    \mathbf{H}_{\rm fs}^{\left( \rm b \right)} [n] = 
    e^{-{\rm j}\frac{2\pi}{\lambda^{\left( \rm b \right)}}\mathbf{D}},
\end{equation}
where 
\begin{equation}
    \mathbf{D} =         
        \begin{bmatrix}
        \mathrm{d}_{1,1} &  \cdots & \mathrm{d}_{1,M_{\rm Tx}}  \\
         \vdots & \ddots   & \vdots  \\
        \mathrm{d}_{M_{\rm Rx},1} & \cdots  & \mathrm{d}_{M_{\rm Rx},M_{\rm Tx}} 
    \end{bmatrix} \in \mathbb{R}^{M_{\rm Rx} \times M_{\rm Tx}}
\end{equation}
represents the distances between specific transmit and receive antenna elements and $\lambda^{\left( \rm b \right)}$ denotes the wavelength.
The Rayleigh channel matrix $\mathbf{H}_{\rm rp}^{\left( \rm b \right)} [n]$ consists of independent complex Gaussian random variables with the power of one.

The path loss coefficient for \ac{MMW} bands is defined by
\begin{equation}
    \eta^{\left( \rm m \right)} = 
    \left( \frac{4 \pi d f_{\rm c}^{\left( \rm m \right)} }{c_0} \right)^2 =
    \left( \frac{4 \pi d \sqrt{\alpha} f_{\rm c}^{\left( \rm s \right)} }{c_0} \right)^2 = 
     \alpha \eta^{\left( \rm s \right)} ,
    \label{eq:fspl_mmWave}
\end{equation}
where $d$ denotes the distance between the transmitter and the receiver, $c_0$ is the speed of light, $\eta^{\left( \rm s \right)}$ represents the free space path loss for the sub-6\,GHz band and $\sqrt{\alpha} = f_{\rm c}^{\left( \rm m \right)} / f_{\rm c}^{\left( \rm s \right)} $ denotes the carrier frequency ratio between the sub-6\,GHz band and the \ac{MMW} band. 
Furthermore, we assume the same transmit power $P_{\rm T}$ for both frequency bands.
Since \ac{MMW} systems tend to use larger bandwidth and can have a noise figure different to sub-6\,GHz systems, we introduce $\beta = \frac{B^{\left( \rm m \right)} F^{\left( \rm m \right)}}{B^{\left( \rm s \right)} F^{\left( \rm s \right)}} $ as the ratio of sub-6\,GHz and \ac{MMW} bandwidths and noise figures.
%Since \ac{MMW} tends to use larger bandwidth than the sub-6\,GHz, we introduce $\beta = B^{\left( \rm m \right)} / B^{\left( \rm s \right)}$ as the ratio of sub-6\,GHz and \ac{MMW} bandwidths.
Then, the obtained pre-beamforming \ac{MMW} \ac{SNR} for the single receive antenna element can be expressed by
\iffalse
\begin{equation}
        \gamma^{\left( \rm m \right)}   
         = \frac{P_{\rm T}}{\eta^{\left( \rm m \right)} k_{\rm B} T B^{\left( \rm m \right)}}
        = \frac{P_{\rm T}}{\alpha \eta^{\left( \rm s \right)} k_{\rm B} T \beta B^{\left( \rm s \right)}}
        = \gamma^{\left( \rm s \right)} \frac{1}{\alpha \beta},   
        \label{eq:snr}
\end{equation}
\fi
\begin{equation}
\begin{split}
         \gamma^{\left( \rm m \right)}   
         = & \frac{P_{\rm T}}{\eta^{\left( \rm m \right)} k_{\rm B} T B^{\left( \rm m \right)} F^{\left( \rm m \right)}}
        = \frac{P_{\rm T}}{\alpha \eta^{\left( \rm s \right)} k_{\rm B} T \beta B^{\left( \rm s \right)} F^{\left( \rm s \right)}} \\
        = & \gamma^{\left( \rm s \right)} \frac{1}{\alpha \beta},   
        \label{eq:snr}
\end{split}
\end{equation}
where $k_{\rm B}$ represents the Boltzmann constant and $T$ denotes the receiver temperature. 
A sub-6\,GHz system with $\alpha \beta$ times higher \ac{SNR} has benefits.
These benefits will be highlighted below.

\subsection{Link Establishment}
Establishing a communication link consists of a training phase and data transmission.
The wireless channel is estimated at sub-6\,GHz and \ac{MMW} frequency bands during the training phase.
The channel estimates obtained at the sub-6\,GHz and \ac{MMW} bands are then utilized for data transmission at the \ac{MMW} band.
We do not transmit data in the sub-6 GHz\,band.
%As the employed sub-6\,GHz bandwidth is typically much smaller than the \ac{MMW} one, data transmission in the sub-6 GHz\,band is not considered.

\subsubsection{Training Phase}
%To make efficient use of the large number of antennas, the transmitter needs to estimate...

% Foundations of MIMO - book
%In \ac{OFDM}, pilots are inserted in the frequency domain occupying $N_p$ subcarriers.
%However, in anticipation of frequency selectivity it is more sensible to intersperse the pilot sequence as uniformly as possible over the $K$ subcarriers.
%Moreover, under the mild premise that the channel be identically distributed, i.e., that all subcarriers look the same statistically, $k_0, \ldots , k_{N_p} − 1$ should be regularly spaced over the $K$ subcarriers, facilitating the interpolation.

\begin{figure}[t]
    \centering
    {\includegraphics[width=\columnwidth]{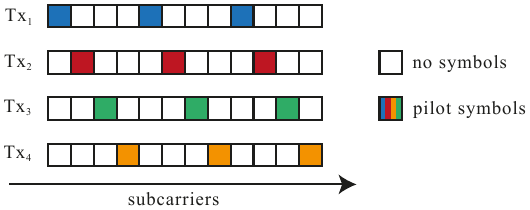}}
    \caption{Pilot symbols at each antenna occupy a certain number of subcarriers such that they do not overlap in the frequency domain.}
    \label{fig:pilot_allocation}
\end{figure}

During the training phase, each transmit antenna is assigned pilot symbols that are known at the receiver. 
The pilot symbols $\boldsymbol{\phi}^{\left( \rm b \right)} [n] \in \mathbb{C}^{M_{\rm Tx} \times 1} $ generated from the \ac{QAM} alphabet are distributed over $N^{\left( \rm b \right)}$ subcarriers so that each antenna occupies a certain number of non-overlapping subcarriers. 
The pilot allocation at the $t$-th transmit antenna is given by
\begin{equation}
    \mathrm{\phi}_t^{\left( \rm b \right)} [n] = 
    \begin{cases}
        \mathrm{\phi}^{\left( \rm b \right)} [n], \quad n \in \{ t, t+M_{\rm Tx}, \ldots, N^{\left( \rm b \right)} \}
        \\
        0, \quad \quad \quad \text{else}
    \end{cases}.
     \label{eq:pilot_allocation}
\end{equation}
A pilot allocation example for $M_{\rm Tx} = 4$ transmit antennas is shown in~\cref{fig:pilot_allocation}.
The input-output relationship for the training phase is given by
\begin{equation} 
    \mathbf{y}^{\left( \rm b \right)} [n] = \mathbf{H}^{\left( \rm b \right)} [n] \boldsymbol{\phi}^{\left( \rm b \right)} [n] + \mathbf{w}^{\left( \rm b \right)} [n],
    \label{eq:training_input_output}
\end{equation}
where the received signal is denoted by $\mathbf{y}^{\left( \rm b \right)} [n] \in \mathbb{C}^{M_{\rm Rx} \times 1} $ and the \ac{AWGN} added at the receiver with the power of $\sigma _w^2$ is denoted by $\mathbf{w}^{\left( \rm b \right)} [n] \in \mathcal{CN}(0,\sigma _w^2 \mathbf{I}_{M_{\rm Rx}})$.
At the receiver, the \ac{LS} estimates of the channel are obtained at pilot positions $n_t \in \{ t, t+M_{\rm Tx}, \ldots, N^{\left( \rm b \right)} \}$.
Further, linear interpolation is performed to estimate the channel at non-occupied subcarrier positions, resulting in an estimated channel $\widetilde{\mathbf{H}}^{\left( \rm b \right)} [n] \in \mathbb{C}^{M_{\rm Rx} \times M_{\rm Tx}}$.
Since the channel is estimated at the receiver, the \ac{CSI} has to be fed back to help the transmitter select the best beamforming matrix.
Therefore, the estimated channel $\widetilde{\mathbf{H}}^{\left( \rm b \right)} [n] \in \mathbb{C}^{M_{\rm Rx} \times M_{\rm Tx}}$ is sent to the transmitter, assuming perfect feedback.
Furthermore, we assume here that the channel in~\cref{eq:channel_model} stays relatively stable over time (long coherence time), so the channel estimation and sending the estimated channel to the transmitter via feedback is only performed infrequently.

\subsubsection{Data Transmission}
%During data transmission, the estimated channel $\widetilde{\mathbf{H}}^{\rm b} [n]$ is then utilized to achieve the highest performance of \ac{MIMO} channel by \ac{SVD} based unitary precoding and combining.
For the data transmission phase, we consider only the mmWave system.
The estimated channels $\widetilde{\mathbf{H}}^{\left( \rm s \right)} [n]$ and $\widetilde{\mathbf{H}}^{\left( \rm m \right)} [n]$ are processed using the three methods proposed in~\cref{sec:methods}.
The resulting \ac{MMW} channel estimate $\overline{\mathbf{H}}^{\left( \rm m \right)} [n]$ is then utilized for achieving the highest performance of the \ac{MIMO} channel by \ac{SVD} based unitary precoding and combining.
%The precoder plays a very important role in \ac{MIMO} communication, enabling a spatial formatting of the transmission on the basis of the \ac{CSI} available at the transmitter.
The compact-form \ac{SVD} of the channel matrix $\overline{\mathbf{H}}^{\left( \rm m \right)} [n]$ can be written as
\begin{equation} 
    \overline{\mathbf{H}}^{\left( \rm m \right)} [n] = 
    \overline{\mathbf{Q}}^{\left( \rm m \right)} [n]  
    \overline{\mathbf{\Sigma}}^{\left( \rm m \right)} [n]
    \left( \overline{\mathbf{F}}^{\left( \rm m \right)} [n] 
    \left( \mathbf{P}^{\left( \rm m \right)} \right)^{1/2}
    \right) ^ {\rm H},
    \label{eq:svd}
\end{equation}
where the semi-unitary matrix $\overline{\mathbf{Q}}^{\left( \rm m \right)} [n] \in \mathbb{C}^{M_{\rm Rx} \times {\ell_{\rm max}}}$ denotes the matrix of left singular vectors and $\overline{\mathbf{F}}^{\left( \rm m \right)} [n] \in \mathbb{C}^{M_{\rm Tx} \times {\ell_{\rm max}}}$ represents the matrix of right singular vectors. 
The singular value matrix is given by
\begin{equation}
    \overline{\mathbf{\Sigma}}^{\left( \rm m \right)} [n] = 
    {\rm diag} \left( \overline{\sigma}_{(1)}^{\left( \rm m \right)} [n], \ldots , \overline{\sigma}_{(\ell_{\rm max})}^{\left( \rm m \right)} [n] \right),
    \label{eq:signular_value_matrix}
\end{equation}
where the $i$-th diagonal element $\overline{\sigma}_{(i)}^{\left( \rm m \right)} [n]$ of the singular value matrix is equals the $i$-th largest singular value of $\overline{\mathbf{H}}^{\left( \rm m \right)} [n]$.
Assuming a full-rank channel, the maximum number of streams is $\ell_{\rm max} = {\rm min} \left( M_{\rm Rx}, M_{\rm Tx} \right)$.
The power loading matrix is given by
\begin{equation}
    \mathbf{P}^{\left( \rm m \right)} = 
    {\rm diag} \left( p_{(1)}^{\left( \rm m \right)}, \ldots , p_{(\ell_{\rm max})}^{\left( \rm m \right)} \right),
    \label{eq:power_matrix}
\end{equation}
where $p_{(1)}^{\left( \rm m \right)} =  p_{(2)}^{\left( \rm m \right)} = \ldots =  p_{(\ell_{\rm max})}^{\left( \rm m \right)} = P_{\rm T} / \ell_{\rm max} > 0$ and the total transmit power constraint
\begin{equation}
    \norm{ 
    \overline{\mathbf{F}}^{\left( \rm m \right)} [n] 
    \left( \mathbf{P}^{\left( \rm m \right)} \right)^{1/2} 
    }_F^2 = P_{\rm T}    
    \label{eq:power_normalization}
\end{equation}
is satisfied by the precoder.

The symbol vector to be transmitted is written as $\mathbf{x}^{\left( \rm m \right)} [n] \in \mathbb{C}^{M_{\rm Tx} \times 1}  $.
Prior to transmission over the wireless channel, the symbol vector $\mathbf{x}^{\left( \rm m \right)} [n]$ is precoded with a precoding matrix $\overline{\mathbf{F}}^{\left( \rm m \right)} [n]$.
At the receiver, combining is performed with a combining matrix $\overline{\mathbf{Q}}^{\left( \rm m \right)} [n]$.
With this notation, the input-output relationship for the data transmission phase is then given by
\begin{equation} 
    \begin{split}
    \mathbf{y}^{\left( \rm m \right)} [n] & =
    \left( \overline{\mathbf{Q}}^{\left( \rm m \right)} [n] \right) ^ {\rm H}
    \mathbf{H}^{\left( \rm m \right)} [n]
    \left( \mathbf{P}^{\left( \rm m \right)} \right)^{1/2}
    \overline{\mathbf{F}}^{\left( \rm m \right)} [n]
    \mathbf{x}^{\left( \rm m \right)} [n] \\
    & + \left( \overline{\mathbf{Q}}^{\left( \rm m \right)} [n] \right) ^ {\rm H}
    \mathbf{w}^{\left( \rm m \right)} [n],        
    \end{split}
    \label{eq:data_input_output}
\end{equation}
where the received signal is denoted by $\mathbf{y}^{\left( \rm m \right)} [n] \in \mathbb{C}^{\ell_{\rm max} \times 1} $ and the \ac{AWGN} added at the \ac{MMW} receiver is denoted by $\mathbf{w}^{\left( \rm m \right)} [n] \in \mathcal{CN}(0,\sigma_w^2 \mathbf{I}_{M_{\rm Rx}})$.

%%%%%%%%%%%%%%%%%%%%%%%%%%%%%%%%%%%%%%%%%

\section{Channel Estimation Methods} \label{sec:methods}
%\iffalse
Due to smaller propagation losses and smaller bandwidth, a sub-6\,GHz system has an $\alpha \beta$ times higher \ac{SNR} compared to the corresponding \ac{MMW} system given the same transmit power (see~\cref{sec:channel_model}). 
This higher \ac{SNR} enables us to achieve more accurate channel estimation for the \ac{MMW} system with the aid of the sub-6\,GHz channel estimate.

We assume that the wavelengths $\lambda^{\left( \rm s \right)}$ and $\lambda^{\left( \rm m \right)}$, as well as the effective distance $\mathbf{D}$ between the transmitter and receiver are known at the receiver, for example, because they are at known positions.
To ensure that the sub-6\,GHz channel estimate occupies the bandwidth $B^{\left( \rm m \right)}$ of the \ac{MMW} system, we first average the $\widetilde{\mathbf{H}}^{\left( \rm s \right)} [n]$ over subcarriers and use the obtained average value to extrapolate the $\widetilde{\mathbf{H}}^{\left( \rm s \right)} [n]$ in the frequency domain.
Next, we take the sub-6\,GHz channel estimate and rotate its phase by $e^{{\rm j} 2\pi \mathbf{D} \xi  }$ as follows
\begin{equation}
    \begin{split}
        & \widehat{\mathbf{H}}^{\left( \rm s \right)}  [n]   
         = \widetilde{\mathbf{H}}^{\left( \rm s \right)} [n] \odot e^{{\rm j} 2\pi \mathbf{D} \xi } \\
        & =  \sqrt{\eta^{\left( \rm s \right)}} \left( \mathrm{A_{\rm fs}^{\left( \rm s \right)}} \widetilde{\mathbf{H}}_{\rm fs}^{\left( \rm s \right)} [n]
        + \mathrm{A_{\rm rp}^{\left( \rm s \right)}} \widetilde{\mathbf{H}}_{\rm rp}^{\left( \rm s \right)} [n] \right)
        \odot e^{{\rm j} 2\pi \mathbf{D} \xi  } \\
        & = \sqrt{\eta^{\left( \rm s \right)}} \left( \mathrm{A_{\rm fs}^{\left( \rm s \right)}} \underbrace{\widetilde{\mathbf{H}}_{\rm fs}^{\left( \rm s \right)} [n] \odot e^{{\rm j} 2\pi \mathbf{D} \xi}  }_{\widetilde{\mathbf{H}}_{\rm fs}^{\left( \rm m \right)} [n] }
        + \mathrm{A_{\rm rp}^{\left( \rm s \right)}} \widetilde{\mathbf{H}}_{\rm rp}^{\left( \rm s \right)} [n] \odot e^{{\rm j} 2\pi \mathbf{D} \xi  } \right)  \\
        & = \sqrt{\eta^{\left( \rm s \right)}} \left( \mathrm{A_{\rm fs}^{\left( \rm s \right)}} \widetilde{\mathbf{H}}_{\rm fs}^{\left( \rm m \right)} [n] 
        + \mathrm{A_{\rm rp}^{\left( \rm s \right)}} \widetilde{\mathbf{H}}_{\rm rp}^{\left( \rm s \right)} [n] 
        \odot e^{{\rm j} 2\pi \mathbf{D} 
        \xi  } \right),
    \end{split}
    \label{eq:phase_shift}
\end{equation}
where $\xi = \frac{1}{\lambda^{\left( \rm s \right)}} - \frac{1}{\lambda^{\left( \rm m \right)}}$. 
Now, the deterministic free-space component of the obtained sub-6\,GHz channel estimate $\widehat{\mathbf{H}}^{\left( \rm s \right)} [n]$ matches the corresponding one of the \ac{MMW} channel estimate $\widetilde{\mathbf{H}}^{\left( \rm m \right)} [n]$.
Introducing the scaling factors $\mathrm{A_{\rm fs}^{\left( \rm s \right)}}$ and $\mathrm{A_{\rm rp}^{\left( \rm s \right)}}$, the $ \widehat{\mathbf{H}}^{\left( \rm s \right)}  [n]$  can be rewritten as
\begin{equation}
    \begin{split}
        \widehat{\mathbf{H}}^{\left( \rm s \right)}  [n]   
         = & \sqrt{\eta^{\left( \rm s \right)}} 
         \sqrt{\frac{0.1 K^{\left( \rm m \right)}}{ 1+ 0.1 K^{\left( \rm m \right)}}}
         \widetilde{\mathbf{H}}_{\rm fs}^{\left( \rm m \right)} [n] \\
        + & \sqrt{\eta^{\left( \rm s \right)}} \sqrt{\frac{1}{1+0.1 K^{\left( \rm m \right)}}}
        \widetilde{\mathbf{H}}_{\rm rp}^{\left( \rm s \right)} [n] 
        \odot e^{{\rm j} 2\pi \mathbf{D} 
        \xi  }.
    \end{split}
\end{equation}
As the \ac{MMW} $K$-factor $K^{\left( \rm m \right)}$ increases, the amplitude of the stochastic Rayleigh component decreases and the phase of the \ac{MMW} channel can be estimated more accurately.

\subsection{Translating}
The most straightforward approach to utilize out-of-band information for establishing the \ac{MMW} link is only to use the phase-rotated sub-6\,GHz channel estimate $\widehat{\mathbf{H}}^{\left( \rm s \right)} [n]$, as follows
\begin{equation}
    \overline{\mathbf{H}}^{\left( \rm m \right)} [n] = \widehat{\mathbf{H}}^{\left( \rm s \right)} [n].
\end{equation}
In its basic implementation, there is no need for any additional calculations, except the phase rotation in~\cref{eq:phase_shift}.
This method performs well for high $K$-factors, where the stochastic Rayleigh component is negligible.
For low $K$-factors, where the stochastic component dominates the deterministic one, there is no benefit to be gained from using this method.

\subsection{Averaging}
To enhance performance for lower $K$-factors, we propose a method that incorporates the phase-rotated sub-6\,GHz channel estimate $\widehat{\mathbf{H}}^{\left( \rm s \right)} [n]$ along with the \ac{MMW} channel estimate $\widetilde{\mathbf{H}}^{\left( \rm m \right)} [n].$
The resulting channel estimate is then given by the average
\begin{equation}
    \overline{\mathbf{H}}^{\left( \rm m \right)} [n] = 
    \frac{\widehat{\mathbf{H}}^{\left( \rm s \right)} [n] +
    \widetilde{\mathbf{H}}^{\left( \rm m \right)} [n]}{2}. 
    \label{eq:avg}
\end{equation}
At moderate $K$-factors, this method yields better results on average than the translating method.
However, this method involves one more implementation step compared to the translating method.
In cases of low and high $K$-factors, there is still room for improvement by assigning weights different to 0.5 to $\widehat{\mathbf{H}}^{\left( \rm s \right)} [n]$ and $ \widetilde{\mathbf{H}}^{\left( \rm m \right)} [n]$ in~\cref{eq:avg}.

\subsection{Weighting}
Performance improvement is achieved through weighting the sub-6\,GHz $\widehat{\mathbf{H}}^{\left( \rm s \right)} [n]$ and \ac{MMW} $\widetilde{\mathbf{H}}^{\left( \rm m \right)} [n]$ channel estimates by a weighting factor $W \left( K^{\left( {\rm m} \right)}, \gamma^{\left( {\rm m} \right)} \right) \in \left[ 0, 1 \right]$ that is a function of the \ac{MMW} $K$-factor $K^{\left( {\rm m} \right)}$ and the \ac{MMW} \ac{SNR} $\gamma^{\left( \rm m \right)}$:
\begin{equation}
    \begin{split}
        \overline{\mathbf{H}}^{\left( {\rm m} \right)} [n] & = 
        W \left( K^{\left( {\rm m} \right)}, \gamma^{\left( \rm m \right)} \right) \widehat{\mathbf{H}}^{\left( {\rm s} \right)} [n] \\ 
        & + \left( 1 - W \left( K^{\left( {\rm m} \right)}, \gamma^{\left( \rm m \right)} \right) \right) \widetilde{\mathbf{H}}^{\left( {\rm m} \right)} [n]
   \end{split}
   \label{eq:weighting}
\end{equation} 
Since the wireless channel is changing, it is important to select an optimum $W \left( K^{\left( {\rm m} \right)}, \gamma^{\left( {\rm m} \right)} \right)$ for each specific combination of $K^{\left( {\rm m} \right)}$ and $\gamma^{\left( \rm m \right)}$.
This ensures that the method is optimal for different wireless channel conditions. 

\begin{figure}[t]
    \centering
    {\includegraphics[width=0.9\columnwidth]{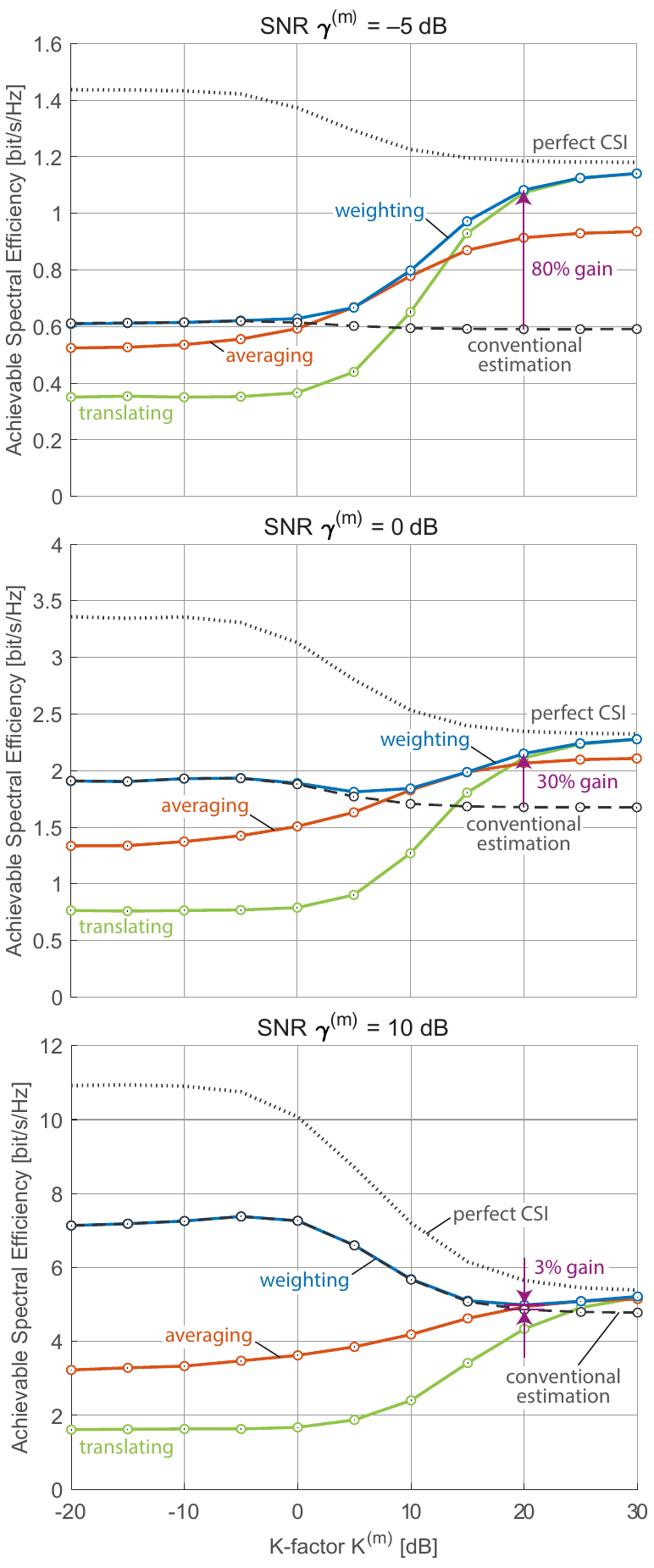}}
    \caption{For the TDL-A 4$\times$4\,\ac{MIMO} channel, the proposed weighting method achieves up to 80\% higher \ac{SE}. The small vertical bars within the circular markers indicate the 95\% confidence intervals.}
    \label{fig:se_tdl}
\end{figure}

During the training phase, we determine and store $W \left( K^{\left( {\rm m} \right)}, \gamma^{\left( {\rm m} \right)} \right)$ for a set of $K^{\left( {\rm m} \right)}$ and $\gamma^{\left( \rm m \right)}$ in a lookup table, see for example~\cref{tab:lookupTable}.
To calculate $W$, we assume perfect knowledge of the \ac{MMW} channel $\mathbf{H}^{\left( \rm m \right)} [n]$, the $ \gamma^{\left( {\rm m} \right)}$ and $K^{\left( {\rm m} \right)}$ by the transmitter and receiver.
Next, we find $W$ that results in the minimum mean channel estimation error
\begin{equation}
    W \left( K^{\left( {\rm m} \right)}, \gamma^{\left( {\rm m} \right)} \right) =
    \operatorname*{argmin}_W 
    E \Bigl\{ W \left( K^{\left( {\rm m} \right)}, \gamma^{\left( {\rm m} \right)} \right) \Bigr\},
\end{equation}
where the mean channel estimation error is defined by
\begin{equation} 
\begin{split}    
    E & \Bigl\{ W \left( K^{\left( {\rm m} \right)},  \gamma^{\left( {\rm m} \right)} \right) \Bigr\} =  \\
    & = \frac{1}{L N^{\left( \rm m \right)}}
    \sum_{l=1}^{L} \sum_{n=1}^{N^{\left( \rm m \right)}} 
    \norm{
    \mathbf{H}_l^{\left( {\rm m} \right)} [n]
    - \overline{\mathbf{H}}_l^{ \left( {\rm m} \right)} [n]
    }_{F}^2.
\end{split}
\label{eq:mse}
\end{equation}
In~\cref{eq:mse}, $\mathbf{H}_l^{\left( {\rm m} \right)} [n]$ denotes the actual channel coefficient and $\overline{\mathbf{H}}_l^{ \left( {\rm m} \right)} [n]$ is the estimated channel coefficient for a specific channel~realization~$l$. 
Results are averaged over    $L = 1000$ different channel realizations.
The lookup table needs to be available at the transmitter and receiver.
During data transmission, the previously calculated $W \left( K^{\left( {\rm m} \right)}, \gamma^{\left( \rm m \right)} \right)$ is taken from the lookup table and used to combine sub-6\,GHz and \ac{MMW} channel estimates.

Compared to the translating method and the averaging method, implementing the weighting method is significantly more demanding. 
It involves an additional phase for populating the lookup table, and requires the lookup table, $\gamma^{\left( \rm m \right)}$ and $K^{\left( {\rm m} \right)}$ to be known at the transmitter and receiver.

\section{Simulation-based Comparison} \label{sec:comparison}
To evaluate the channel estimation methods proposed, we simulate the achievable \ac{SE} in a frequency-selective channel.
The parameters of the simulation are summarized in~\cref{tab:simParams}.
For the frequency-selective channel model, we utilize the \ac{TDL-A} model with the deterministic channel component for the urban macro (UMa) scenario from~\cite{3gpp.38.901}.
Corresponding \ac{RMS} delay spread values for the case without the deterministic channel component are given in~\cref{tab:simParams}.
The weighting factor lookup table for the \ac{TDL-A} channel model is given in~\cref{tab:lookupTable}. 

We consider the achievable \ac{SE} (achievable rate per bandwidth) as the main performance metric for \ac{MIMO} systems.
Since \ac{SE} is rate per bandwidth, this metric does not depend on the employed transmission bandwidth and therefore facilitates direct comparison of results.
The achievable \ac{SE} in bits$/$s$/$Hz averaged over $N^{\left( \rm m \right)}$ subcarriers is given by
\begin{equation} 
    \mathrm{SE}= 
    \frac{1}{N^{\left( \rm m \right)}} \sum_{n=1}^{N^{\left( \rm m \right)}} 
    \log_2 \left( 1 + \mathrm{SINR} [n] \right)
    \label{eq:se}
\end{equation}
with 
\begin{equation} 
    \mathrm{SINR} [n] = \frac{ 
    \sum\limits_{ \substack{\mu=1 \\ \nu = \mu}}^{\ell_{\rm max}} 
    \left| \overline{\mathrm{G}}_{\rm \mu,\nu}^{\left( {\rm m} \right)} [n] \right|^2 }
    { \sum\limits_{\mu=1}^{\ell_{\rm max}} 
    \sum\limits_{\substack{\nu=1 \\ \nu \neq \mu}}^{\ell_{\rm max}} 
    \left| \overline{\mathrm{G}}_{\rm \mu, \nu}^{\left( {\rm m} \right)} [n] \right|^2  
    + \sigma _w^2
    \norm{\overline{\mathbf{Q}}^{\left( {\rm m} \right)} [n] }^2_F}.
%    \left| \overline{\mathrm{Q}}_{\rm \mu, \nu}^{\left( {\rm m} \right)} [n] \right|^2  }. 
    \label{eq:sinr}
\end{equation}
In~\cref{eq:sinr}, $\overline{\mathrm{G}}_{\rm \mu,\nu}^{\left( {\rm m} \right)} [n]$ with $\mu, \nu \in \{ 1, \ldots, \ell_{\rm max} \} $ represent entries of the channel gain matrix $\mathbf{G}^{\left( {\rm m} \right)} [n] \in \mathbb{C}^{\ell_{\rm max} \times \ell_{\rm max}}$ for $n$-th subcarrier which is given by
\begin{equation} 
    \overline{\mathbf{G}}^{\left( {\rm m} \right)} [n]= 
    \left( \overline{\mathbf{Q}}^{\left( {\rm m} \right)} [n] \right) ^ {\rm H} 
    \mathbf{H}^{\left( {\rm m} \right)} [n]
    \left( \mathbf{P}^{\left( \rm m \right)} \right)^{1/2}
    \overline{\mathbf{F}}^{\left( {\rm m} \right)} [n] .
    \label{eq:chgain}
\end{equation}

After averaging over $L=1000$ different channel realizations, we obtain the simulation results shown in~\cref{fig:se_tdl}.
As a conventional channel estimation method, we use the achievable \ac{SE} for the case that only the \ac{MMW} channel estimate $\overline{\mathbf{H}}^{\left( {\rm m} \right)} [n] = \widetilde{\mathbf{H}}^{\left( {\rm m} \right)} [n] $ is employed.
The achievable \ac{SE} for the case of perfect \ac{CSI} ($\overline{\mathbf{H}}^{\left( {\rm m} \right)} [n] = \mathbf{H}^{\left( {\rm m} \right)} [n]$) is shown in~\cref{fig:se_tdl} as well. 

\begin{table}[t]
    \centering
    \caption{Simulation Parameters}
    \label{tab:simParams}
    \begin{tabular}{rcc}
        \hline
        \textbf{Parameter}                          & \multicolumn{2}{c}{\textbf{Value}} \\ \hline
        Frequency Band                              & sub-6 GHz         & mmWave         \\
        Carrier Frequency $f_{\rm c}$  [GHz]        & 2.55              & 25.5           \\
        Wavelength $\lambda$ [cm]                   & 11.76             & 1.176          \\
        Bandwidth $B$ [MHz]                         & 10.08             & 100.8            \\
%        Number of Subcarriers $N$                   & 168               & 1680              \\
%        Sampling Rate $f_{\rm s}$ [MHz]             & 20.16             & 201.6              \\
        Subcarrier Spacing $\bigtriangleup f$ [kHz] & 60                & 60             \\
        Cyclic Prefix $t_{\rm CP}$ [$\mu$s]         & 1.19              & 1.19             \\
        Noise Figure $F$ [dB]                       & 3                 & 3            \\        
        Antenna Spacing [cm]                        & 0.05\,$\lambda$    & 0.5\,$\lambda$   \\
        Channel Model                               & TDL-A 1148\,ns    & TDL-A 841\,ns   \\ \hline
    \end{tabular}
\end{table}

\textbf{Results at \ac{SNR} of $-$5\,dB:} When the translating method is used at a high $K$-factor $K^{\left( {\rm m} \right)}$, it leads to $80\%$ improvement in achievable \ac{SE} as compared to the case when estimating using only the \ac{MMW} band.
However, the performance of the translating method is poor at low $K$-factor.
The averaging method yields better results at low $K$-factor.
At high $K$-factor, the averaging method performs worse than the translating method.
Ultimately, the weighting method is for all $K$-factors at least as good or better than the translating method and the averaging method.

%By selecting the appropriate weight $W \left( K, \gamma^{\left( {\rm m} \right)} \right)$ from~\cref{tab:lookupTable}, the sub-6\,GHz and \ac{MMW} channel estimates are combined such that the achievable \ac{SE} is greater than or equal to the reference value for each \ac{SNR} and $K^{\left( {\rm m} \right)}$. %when only the mmWave estimate is used.
%For low $K$-factors, a low $\alpha$ is chosen so that the \ac{MMW} portion dominates, while for high K factors, the reverse is true.
%It is important to note that this method also selects the correct $\alpha$ in the transition region.

\textbf{Results at \ac{SNR} of 0\,dB:}
Again, the weighting method outperforms the translating method and the averaging method, although the gain compared to conventional channel estimation at high $K$-factors is less (30\%).

\textbf{Results at \ac{SNR} of 10\,dB:}
The gain compared to conventional channel estimation at high $K$-factors is now only 3\%.
Therefore, it is questionable if the effort justifies the gain.

%With an \ac{SNR} of $\gamma^{\left( \rm m \right) }=-$5\,dB, even a low $K$-factor is enough to observe an improvement in achievable \ac{SE} when using sub-6\,GHz out-of-band information.
%More precisely, if the $K^{\left( {\rm m} \right)}$ is greater than 0\,dB, the \ac{SE} can be improved by up to around 80\%.
%When the \ac{SNR} is $\gamma^{\left( \rm m \right) }=$0\,dB, the $K^{\left( {\rm m} \right)}$ value from which using sub-6\,GHz out-of-band information provides an improvement is shifted to around 5\,dB.
%In that case, the achievable \ac{SE} can be improved up to around 30\%.
%At a high \ac{SNR} of $\gamma^{\left( \rm m \right) }=$10\,dB, the enhancement in achievable \ac{SE} is minor, and it can only be increased by around 3\%.

%The SE is obtained by...
%The SE is then calculated by obtaining...
%We observe...

%Figure X shows the sum throughput of both UEs for the three different schedulers over the SNR. 
%X and Y perform similar. 
%The method X gains about 1.8−2 compared to the...
%Figure X shows a comparison of the SEs of the ... when different channel estimation methods are employed.
%There is a slight difference in the SE performance of X and Y, because....
%The reason for this is that...
%The SE of X is lower than that of our... because it underestimates...

\section{Conclusion} \label{sec:conclusion}
In this paper, we propose three novel channel estimation methods for \ac{MMW} \ac{MIMO} systems based on using sub-6\,GHz out-of-band information.
The \ac{SNR} and $K$-factor affect the performance of the proposed channel estimation methods to a great extent.
The proposed methods lead to a significant increase in the achievable \ac{SE} in the low-\ac{SNR} regime, while the gains are negligible in the high-\ac{SNR} regime.
The translating method performs well at high $K$-factors.
The averaging method performs well at moderate $K$-factor.
The weighting method is better (high $K$-factor) or at least as good (low $K$-factor) as the conventional channel estimation method using only the \ac{MMW} band.

% \begin{itemize}
%     \item The translating method performs well at high $K$-factors.
%     \item The averaging method performs well at moderate $K$-factor.
%     \item Simulation results show that the proposed weighting method is better (high $K$-factor) or at least as good (low $K$-factor) as the conventional channel estimation method using only the \ac{MMW} band.
% \end{itemize}

\begin{table}[t]
\centering
\caption{Weighting Factor W Lookup Table \\ TDL-A Channel}
\label{tab:lookupTable}
\begin{tabular}{l|llllllll}
\multicolumn{1}{c|}{\multirow{2}{*}{$K^{\left( {\rm m} \right)}$\,[dB]}} & \multicolumn{8}{c}{$\gamma^{\left( {\rm m} \right)}$ [dB]}                      \\
\multicolumn{1}{c|}{}                                   & -15  & -10  & -5   & 0    & 5    & 10   & 15   & 20   \\ \hline
-20                                                     & 0.83 & 0.92 & 0.98 & 1    & 1    & 1    & 1    & 1    \\
-15                                                     & 0.81 & 0.91 & 0.97 & 0.99 & 1    & 1    & 1    & 1    \\
-10                                                     & 0.76 & 0.86 & 0.95 & 0.99 & 1    & 1    & 1    & 1    \\
-5                                                      & 0.63 & 0.75 & 0.89 & 0.97 & 1    & 1    & 1    & 1    \\
0                                                       & 0.4  & 0.55 & 0.76 & 0.89 & 0.99 & 1    & 1    & 1    \\
5                                                       & 0.19 & 0.33 & 0.56 & 0.79 & 0.93 & 1    & 1    & 1    \\
10                                                      & 0.1  & 0.18 & 0.35 & 0.61 & 0.85 & 0.99 & 1    & 1    \\
15                                                      & 0.04 & 0.1  & 0.19 & 0.39 & 0.66 & 0.87 & 0.99 & 1    \\
20                                                      & 0.01 & 0.03 & 0.1  & 0.2  & 0.41 & 0.68 & 0.87 & 0.99 \\
25                                                      & 0    & 0    & 0.02 & 0.1  & 0.2  & 0.41 & 0.68 & 0.87 \\
30                                                      & 0    & 0    & 0    & 0.02 & 0.1  & 0.2  & 0.41 & 0.68
\end{tabular} \\[1em]
\begin{flushleft}Note that for high $K^{\left( {\rm m} \right)}$, $W$ is close to 1, as the phase-rotated sub-6\,GHz estimate significantly improves channel estimation. Conversely, for low $K^{\left( {\rm m} \right)}$, the phase-rotated sub-6\,GHz estimate does not play a significant role. In this case, $W$ is close to 0. 
\end{flushleft}
\end{table}

%proposed methods lead to significant improvements in terms of achievable \ac{SE}, especially in the low-\ac{SNR} regime and at higher $K^{\left( {\rm m} \right)}$-factors.
%On the other hand, we observe that in the high-\ac{SNR} regime, the methods provide only minimal improvements.

%In this contribution, we analyze ...
%We compare three different ...
%The presented results show that ...

\section*{Acknowledgment}
This work was supported by the Austrian Research Promotion Agency (FFG) via the research project Intelligent Intersection (ICT of the Future, Grant 880830).
The work of M. Hofer and T. Zemen was supported by the principal scientist grant DEDICATE.
The work of M. Mussbah was supported by the Christian
Doppler Research Association.

% \IEEEtriggeratref{7}
\bibliography{references}
\bibliographystyle{IEEEtran}

\end{document}